\documentclass[conference]{IEEEtran}

\usepackage{amsmath}
\usepackage{amssymb}

\usepackage{color}
\usepackage{dcolumn}
\usepackage{float}
\usepackage{graphicx}
\usepackage[latin9]{inputenc}
\usepackage{rotating}
\usepackage{psfrag}
\usepackage{tabularx}
\usepackage[hyphens]{url}
\usepackage{fancyvrb}

\usepackage[bookmarks, bookmarksopen, bookmarksnumbered]{hyperref}
\usepackage[all]{hypcap}
\newcommand{\simfactory}{SimFactory}

\newcommand{\mdb}{{Machine Database}}

\newcommand{\udb}{{User Database}}

\newcommand{\ini}{{INI}}
\newcommand{\blockstart}{{\textless}{\textless}}
\newcommand{\negfigurespace}{\vspace{-2mm}}
\begin{document}

\title{Simulation Factory: Taming Application Configuration and
  Workflow on High-End Resources}

\author{
  \IEEEauthorblockN{Michael W. Thomas}
  \IEEEauthorblockA{Center for Computation \& Technology/ \\
    Department of Computer Science \\
    Louisiana State University}
%    Baton Rouge, LA 70803, USA}
  \and
  \IEEEauthorblockN{Erik Schnetter}
  \IEEEauthorblockA{Center for Computation \& Technology/ \\
    Department of Physics \& Astronomy \\
    Louisiana State University}
%    Baton Rouge, LA 70803, USA}
}

\maketitle

\begin{abstract}
  Computational Science on large high performance computing resources
  is hampered by the complexity of these systems.  Much of this
  complexity is due to low-level details on these resources that are
  exposed to the application and the end user.  This includes (but is
  not limited to) mechanisms for remote access, configuring and
  building applications from source code, and managing simulations and
  their output files via batch queue systems.
  
  These challenges multiply in a modern research environment, where a
  research collaboration spans multiple groups, often in loosely
  defined international collaborations, where there is a constant
  influx of new students into multi-year projects, and where
  simulations are performed on several different resources.
  
  The \emph{Simulation Factory} addresses these challenges by
  significantly simplifying remote access, building executables, and
  managing simulations.  By abstracting out the low-level differences
  between different resources, it offers a uniform interface to these
  resources.  At the same time, it can enforce certain standards for
  performing simulations that encapsulate best practices from
  experienced users.  Furthermore, \simfactory's automation avoids
  many possible user errors that can in the worst case render
  month-long simulations worthless.
  
  The Simulation Factory is freely available under an open source
  license.
\end{abstract}

\maketitle

%\setcounter{tocdepth}{3}
%\tableofcontents \newpage

%%%%%%%%%%%%%%%%%%%%%%%%%%%%%%%%%%%%%%%%%%%%%%%%%%%%%%%%%%%%%%%%%%%%%%%%%%%%%%%%

\section{Introduction}

% {ES: Review 2 suggest to compare to ClayWorks.  ClayWorks
%   addresses a very different problem and uses a very different
%   approach; I believe it is off-topic and will ignore ClayWorks here.}

%\todo{ES: Review 4 requests to define all terms at their first usage.
%  Read paper and check.  Michael, can you do this?}

% {ES: Review 4 complains that we refer to figures ``above''.
%  Ensure that ``above'' figures really come before the citation.
%  Michael, can you do this?}

Although the speed and performance of high end computers have
increased dramatically over the last decade, the ease of using such
parallel computers has not progressed.  The time and effort required
to develop and deploy parallel codes and to manage and post-process
simulations has become a bottleneck in many areas of science and
engineering.  The difficulty of using high performance computing is
recognized as one of the most significant challenges today in many
areas of science and engineering.  Simplifying this is
crucial to expanding the understanding in these fields via the use of
high performance computing.

To simplify code development, large software projects use a modular
design or may employ software frameworks.  These aid in handling
different versions, managing many users, distributed code development,
and complex machine architectures.
\emph{Cactus} \cite{Goodale02a, cactusweb1} is one such software
framework for science applications which is used to simulate physical systems in
different fields of science such as black holes and neutron stars in
general relativity.  As in other software frameworks, applications are
built from separately developed and tested components.

To effectively begin using high performance computing (HPC), one has to
overcome a set of unique technical challenges which have their roots
in the fact that HPC systems are expensive and unique, very different
from commodity workstations.  These challenges include:
\begin{itemize}
\item \emph{preventing lock-in:} the fundamental need to use multiple
  HPC systems, since individual systems may be unavailable at any
  given time, and since HPC systems have a much shorter life time than
  software;
\item \emph{lack of standards:} the fact that each HPC system is often
  designed independently and can have very different hardware
  architectures and usage policies;
\item \emph{expensive mistakes:} the low-level interface that HPC
  systems offer to manage simulations which require batch
  submissions and frequent maintenance of ongoing simulations with dire
  consequences for user errors that can destroy weeks of data.
  % {ES: Review 4 requests a motivation and an example here.}
  For example, a job may accidentally be started in the wrong
  directory and then overwrite existing data, or data may accidentally
  be deleted before they are successfully archived.  Such errors
  happen ``naturally'' if a user has to manually manage several jobs
  at the same time.
\end{itemize}
For example, HPC systems differ in the available software and their
versions, their directory structure and file systems, queuing systems,
and policies, in addition to their hardware differences.  This makes
porting codes, handling data, and setting up simulations very tedious
tasks on each system anew.  Yet for scientific results none of these
differences actually matter.
Because HPC systems have a relatively small number
of users compared to commodity systems, there is no obvious economic 
motivation to improve the situation, since most heavy users of HPC systems
have already become adept at handling these differences.
% {ES: Review 4 didn't understand the ``competing'' part.}
At the same time, large HPC centers compete for funding (e.g.\ the 11
sites providing TeraGrid resources for NSF) and for users.  They
thus have to ensure they stand out and don't merely copy another
centre's practice.  This makes standardization difficult.

Due to these challenges, users typically adopt one of two modes of
operation: one, where one decides on one particular version of the
software and one HPC system, and then uses these exclusively, and
another, where one leaves these details to graduate students who then
have to spend a significant amount of time with these frustrating
low-level details. 
These challenges also lead to the perception that high performance
computing is very difficult and hinders update by newcomers.

To help drive forward a broader vision of HPC, we argue that handling
these details should not be the responsibility of users in the first
place, but should rather be folded into the actual systems.  We
introduce the \emph{Simulation Factory}, a high-level interface to
managing source code, accessing remote systems, and performing
simulations.  The Simulation Factory, also known as \emph{SimFactory}, offers an abstract work flow
covering these tasks and offers a small set of commands for the
most-often required actions.  All system details are described in a
\emph{machine database}.  Thus, the Simulation Factory offers a uniform
interface to users and renders different HPC system into fungible~\footnote{Fungibility - A good or commodity capable of mutual substitution: \url{http://en.wikipedia.org/wiki/Fungibility}}
resources.

\subsection{Related Work} 
\label{sec:related}

% {ES: Review 3 suggests to compare to ``distributed configuration
%   management'' in the literature; point out that we are using HPC
%   systems where we do not have root access.  Suggested systems for
%   configuration management are: SmartFrog, Autonomia.  For deploying
%   and execution: BioOpera/Jopera, ProActive.}
% {ES: SmartFrog: requires a Java runtime engine -- seems to update
%   components at run time, and requires to be installed; this is
%   different from our focus.}
% {ES: Similarly, Autonomia is a large system that needs to be
%   installed, and which does not target HPC systems.}
% {ES: BioOpera is dead. JOpera is for workflow management; this
%   doesn't apply for us.}
% {ES: ProActive is difficult to google.}

Many component frameworks face problems of collecting components from
various independent sources.  The Eclipse~\cite{eclipseweb} approach is
probably most similar to the one we describe below, with the exception
that installing or updating Eclipse components requires Eclipse (and a
graphical user interface) to be running, which can be inconvenient or
prohibitive on HPC systems.  Eclipse also automatically enforces
dependencies between different versions of components; this is
convenient or necessary for a framework where most users are only
end-users, but probably inconvenient for a scenario where most users
are also developers and modify code so that the concept of a
``version'' with a well-defined API is not well defined.

UBIQIS (ubiquitous install)~\cite{ubiqisweb} is a system for
automatically fetching and installing software from the web or even a
peer to peer network.  It provides a place from where one can
automatically get software, but it does not require root permissions
to install, and the software repository is not centralized.  It
automatically caches requests for software, automatically fetching
dependencies as they are needed.  We are exploring in how far UBIQIS
can be incorporated into the Simulation Factory \cite{ES-Brandt2009b}.

Since its first release, Cactus has included a mechanism for groups or individuals to maintain their own set of configuration files for compiling their 
particular applications. A default set of files was also distributed along side Cactus which were sufficient to build the compete set of distributed thorns
on machines supported by the Cactus group with different option files provided for different purposes (e.g.\ single processor, MPI, debug). Since these option files followed a particular naming convention and were available anonymously via CVS they could be included in higher level tools for 
Cactus.

For automated configuring and building, we describe and compare to related
approaches in section \ref{sec:configuring}, where we also discuss the
requirements that these have to satisfy.

An early driver for automated processes for compilation and run time support came from Cactus' role in the development and prototyping
of Grid Computing scenarios for scientific computing. For example, in
2002 and 2003, a set of tools (dubbed ``GridTools''~\cite{GridTools}, and written in Perl) was developed by Ian Kelley that acted as a compilation, 
deployment, testing and prototyping infrastructure for Cactus on top of the Globus Toolkit~\cite{globusweb} command-line tools, which are a set 
of tools for grid computing that provide authentication, file transfer, job submission, and more.

% {ES: Review 5 suggest to introduce the Globus Toolkit where it is
%  first mentioned, and use a consistent name for it everywhere it is
%  mentioned.  Michael, can you do this?}

GridTools gave users the added benefit of being able to easily run predefined tests upon a set of resources.  
GridTools could do such things like query Globus Toolkit gatekeepers or information services or perform GSI-ftp transfer tests, and then aggregate all this information into a common view that showed  which parts of our testbed infrastructure were not performing as expected.  
Since GridTools was created using a modular approach for test integration, with most of its core functionality stored in module libraries, 
it was very trivial to add new or manipulate pre-existing tests.  Leading up to the HPC Challenge Awards at SC~2002 GridTools was extensively used
 to test, setup, and verify our large testbed infrastructure which included over 80 different production HPC resources and 
 subsequently won the HPC Challenge Award for ``Most Geographically Distributed Applications and Most Heterogeneous Set of Platforms.''  

Mock et al.~\cite{Mock02} describe a batch script generator that was used
successfully from SDSC with Cactus, connecting to about 80 different
machines for a demonstration performed at SC~2001.
This script generator runs as a web application, not necessarily tied
into portal, to generate all the batch scripts needed for the
TeraGrid machines used during this demonstration.

% {ES:
%   Review 6 suggest to refer to the ``And away we go'' which introduced
%   the idea of an mdb.}
Yoon et al.~\cite{Yoon05} describe the difficulties in launching
applications on HPC systems.  Their approach to solving this problem
differs in two crucial aspects from the approach described here:
First, it relies on web services instead of on tools that are
already available on standard HPC systems which makes it difficult to quickly
deploy their solution on a new HPC system.  Second, the descriptions
of the available resources (the machine database) are maintained in a
distributed manner by the resource owners.  This is highly problematic
because these descriptions are often a low priority since the majority
of users do not need them. As a consequence, errors in resource 
descriptions may not be corrected in a timely manner, rendering resources unusable. 

%This is highly
%problematic because resource owners may maintain these descriptions
%only with a low priority, since the majority of users do not (yet?)
%need them.  As a consequence, errors in resource descriptions may not be
%corrected in a timely manner, rendering resources unusable.

% {ES: Michael, this paragraph is new. Please proof-read it and
%   check that it blends with its neighbourhood.}

% {ES: Review 1 suggests to describe more related work, and to
%   emphasise what makes SimFactory unique.  I think that using (rsync)
%   is not unique; the important main idea is to describe remote access
%   (push model) and remote location (directory) in the mdb, which must
%   be similar to what enterprise software management systems offer as
%   well.}
% {ES: We did add more related work, and we describe better how SF is
%   unique.}

\vspace{3ex}\noindent
This paper is structured as follows.  In section \ref{sec:cactus} we
introduce the Cactus Software Framework as model for the kind of
applications that our infrastructure supports.  In section
\ref{sec:simfactorybasic} we introduce the Simulation Factory and
describe its basic concepts, and elaborate on some implementation
issues in section \ref{sec:implementation}.  We describe our future
plans in section \ref{sec:future}.

%%%%%%%%%%%%%%%%%%%%%%%%%%%%%%%%%%%%%%%%%%%%%%%%%%%%%%%%%%%%%%%%%%%%%%%%%%%%%%%%

\section{Application Example: Cactus Software Framework}
\label{sec:cactus}

Cactus \cite{Goodale02a, cactusweb1} is a software framework for
science applications which is used to simulate physical systems in
many fields of science and engineering such as black holes and
neutron stars in general relativity.  As in other software frameworks,
applications are built from separately developed and tested
components.  Cactus is an open-source, modular, and portable
programming environment for collaborative high performance computing
(HPC\@).  It was designed and written specifically to enable
scientists and engineers to develop and perform the large-scale
simulations needed for modern scientific discovery across a broad
range of disciplines.

The Cactus code base is structured as a central part, called the
\emph{flesh} that provides core routines, and components, called
\emph{thorns}.  The flesh is independent of all thorns and provides
the main program, which parses the parameters and activates the
appropriate thorns, passing control to thorns as required.  By itself,
the flesh does not do any science; to do any computational task the
user must compile in thorns and activate them at runtime.

A thorn is the basic working component within Cactus.  All
user-supplied code goes into thorns, which are, by and large,
independent of each other.  Thorns communicate with each other via
calls to the flesh API or, more rarely, custom APIs of other thorns.
The Cactus component model is based upon tightly coupled subroutines
working successively on the same data, although recent changes have
broadened this to allow some element of spatial workflow.  The
connection from a thorn to the flesh or to other thorns is specified
in configuration files that are parsed at compile time and used to
generate glue code that encapsulates the external appearance of a
thorn.  At runtime, the executable reads a parameter file that details
which thorns are to be active and specifies values for the control
parameters for these thorns.

% User thorns are generally stateless entities; they operate only on
% data which are passed to them.  The data flow is managed by the flesh.
% This makes for a very robust model where thorns can be tested and
% validated independently, and can be combined at run-time in the manner
% of a functional programming language.  Furthermore, thorns contain
% test cases for unit testing.  Parallelism, communication, load
% balancing, memory management, and I/O are handled by a special
% component called \emph{driver} which is not part of the flesh and
% which can be easily replaced.  The flesh (and the driver) have
% complete knowledge about the state of the application, allowing
% inspection and introspection through generic APIs.

The current version of Cactus provides many computational modules for
finite difference based methods and has been very successful as
indicated by the large number of scientific publications it has
enabled.  There exist currently (July 2010) more than 500 thorns in
over 50 arrangements at various sites world-wide, many of which are
publicly available.
An international consortium has recently (June 2010) released
the Einstein Toolkit~\cite{einsteintoolkitweb}, a complete,
production-level, open-source
set of components for relativistic astrophysics simulations that uses
Cactus framework.  We describe some details of the component structure
of the Einstein Toolkit in \cite{ES-Schnetter2008n}.

\begin{figure}
  \includegraphics[width=\linewidth]{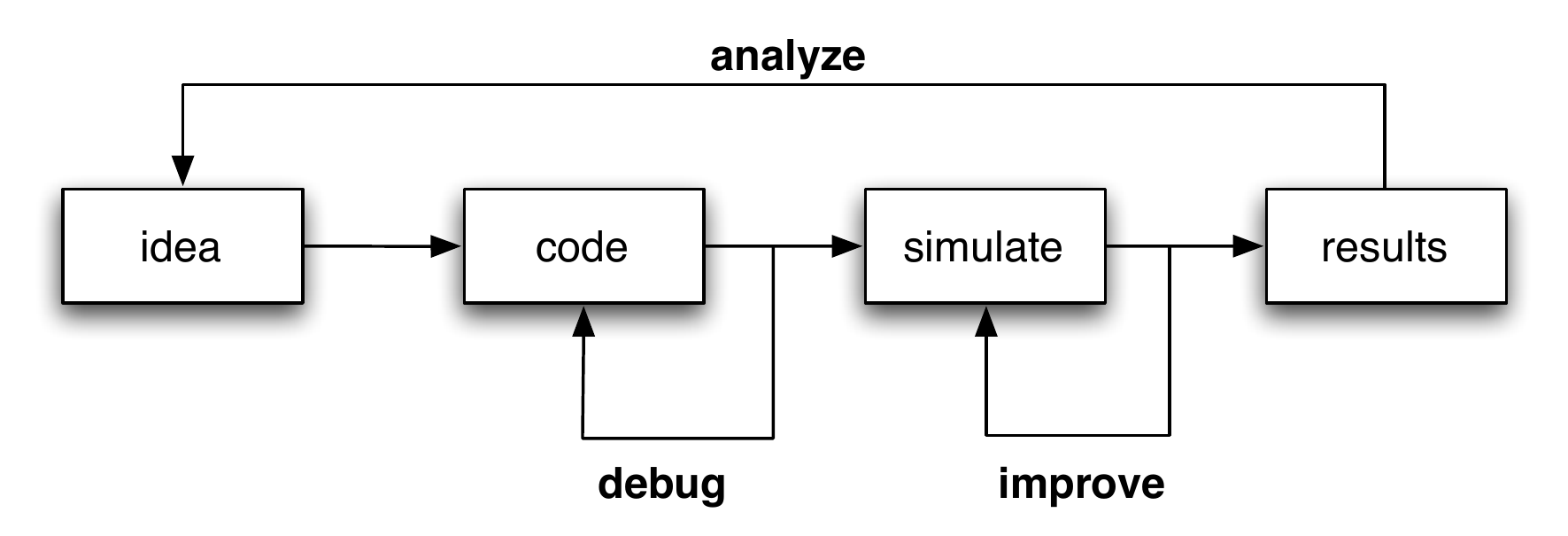}
  \caption{General life cycle of a simulation, beginning with an idea,
    proceeding to a code development stage and then a production
    simulation stage, and ending with simulation results being
    post-processed and analyzed.}
  \label{fig:lifecycle}
  \negfigurespace
  \negfigurespace
\end{figure}

Most researchers who use Cactus are developing their own code,
combining both public and self-written components.  This mode of work,
whereas an existing code base is easily augmented by new modules, is
made possible by Cactus's component structure which does not require
nor permit any kind of centralized control.  Researchers working in
this way then need to build and run their code on various platforms.
We depict the resulting life cycle of a simulation science project in
figure \ref{fig:lifecycle}.

% {ES: Review 4 calls this paragraph redundant.}
% Building any kind of real-world code on today's HPC systems is a
% difficult undertaking, because these systems differ in many respects,
% such as hardware architecture, software installation, and site
% policies.  This requires
% the user to carefully study the documentation of each HPC system that
% he/she wants to use, and modify makefiles and build parameters
% correspondingly.  This is a complex task that requires knowledge of
% arcane parts of the Unix operating system, and is usually only
% possible for experts.  Cactus, by itself, is no different in this
% respect.

As an interesting side note, we want to remark that we have a sizable
number of computer science researchers who are using Cactus.  These
researchers are often not familiar with the day-to-day usage of HPC
systems since they have never been trained in the corresponding
low-level details, and thus cannot undertake research in real-world
systems.  Nevertheless, these researchers are important for developing
new capabilities with a sound theoretical backing.  It is often
difficult, if not impossible, to close the gap between a prototype and
a real-world implementation.

Using HPC systems to perform simulations is similarly a complex task.
These systems are run in \emph{batch mode}, where one has to prepare a
shell script that runs the actual code, and then submit this script to
a batch system.  After some time (hours or days), the script is
executed, and its output and error messages are returned to the user.
It goes without saying that debugging such scripts is very time
consuming due to the long turn-around time and any user error can
delay a simulation project by several days.

To add insult to injury, the maximum possible run times in such
queuing systems are usually measured in hours, and not more than two
days at best.  This needs to be compared to production-level
simulations that can require run times of the order of
weeks.  To deal
with this, a simulation has to \emph{checkpoint} itself before its
queue time runs out and then \emph{restart} when it receives another
time slot in the batch system.
% {ES: Review 6 complains that we claim generality (``all HPC
%   systems''), which we need to back up by examples.  ``For the
%   systems, it is simply not true all HPC systems do not allow long
%   running jobs -- that is exactly how the largest systems are used, to
%   run the biggest jobs until completion on large numbers of
%   processors.''}
In addition, if one uses a large number of nodes, there is an
increased chance of system failures that will also require
checkpointing and restarting.
Automating the corresponding frequent
re-submissions to the queuing systems is called \emph{job chaining} or
\emph{presubmission}.  Dealing with queuing systems can be very
stressful for human beings because even small errors can invalidate
or destroy weeks of simulation results.  Each simulation day may
consume 10k to 100k of CPU hours, corresponding to \$1k to \$10k in US
currency (if bought on the free market).

The problems described here exist on all commonly available HPC
systems and exist for virtually all current HPC applications
(including Cactus).  It is clear that this situation is highly
unsatisfactory and that there is a dire need to shield the user from
system-level details and low-level simulation maintenance operations.

Once a researcher has moved from postdoc to faculty, he or she usually
does not have the time any more to use HPC systems daily.  Once this
practice is gone, it is very difficult to get ``back into the
groove'', leading to highly unsatisfactory situations where faculty
(who often originally developed the software!) cannot use them any
more.
% It is clear that this gap leads to tensions in many research groups.

%%%%%%%%%%%%%%%%%%%%%%%%%%%%%%%%%%%%%%%%%%%%%%%%%%%%%%%%%%%%%%%%%%%%%%%%%%%%%%%%

\section{Simulation Factory: Basic Concepts}
\label{sec:simfactorybasic}

\simfactory\ evolved as an application to manage the idiosyncrasies
present in HPC systems.  The complexity found herein is a result of
the uniqueness of each of these machines and in the uniqueness of the
codes users deploy there.  Real-world input from these users allowed
\simfactory\ to move beyond a prototype and into a practical,
pragmatic application that truly aids non-computer-science research.
While the design of \simfactory\ may seem overly complex in places,
this only reflects the complexity that exists in the world of high
performance computing.

\simfactory\ addresses three main goals: \emph{source code management}, \emph{configuring and building simulations}, and 
\emph{managing and executing simulations}. Building on a \mdb, \simfactory\ can accomplish these goals consistently 
and transparently across any pre-configured HPC resource. Each of the
three main goals contains unique challenges that we describe below. Figure~\ref{fig:simworkflow} below describes the lifecycle
of a simulation as it directly relates to simfactory commands.

\subsection{Managing Source Code}

Many research codes are developed in an open and/or collaborative
manner, where end users not only use applications, but also make
modifications or add their own modules.
This is in particular facilitated by the component
model of the Cactus framework (see section \ref{sec:cactus} above)
that we use as model application here.
A user may start with a certain version of an application code, make
these modification on a workstation or notebook, and then move on to an HPC
resource after some testing.

While doing so, it is important to keep track of the different
versions and modifications of the application source tree.  It is
usually not possible to use version control systems at this stage,
because (a) the user may lack direct write permissions to the
repository, or (b) the repository rules accept only well-tested and
finished changes.  To facilitate development that spans multiple
machines and testing on different platforms, \simfactory\ offers the
following model:
\begin{itemize}
\item the user chooses a single \emph{home system} where he/she
  performs all code development;
\item this home system contains the \emph{authoritative version} of the
  source code, or contains multiple versions in different directories;
\item all other systems are set up as \emph{mirrors} of the home
  system, i.e.\ usually the source trees on these systems is not
  directly edited;
\item \simfactory\ provides a convenient (easy and quick) method to
  \emph{synchronize} (mirror) source trees.
\end{itemize}
In particular, this implies that one accesses code repositories (cvs,
svn, git, \ldots) only from the home system.
% {ES: Review 1 is
%  confused about this: ``More technically, with respect to Section
%  III.A \& B, I am not sure to understand how SimFactory deals with
%  source codes owned by several groups: does a user need to set up a
%  global repository for all codes or is it possible to mirror external
%  codes to her personal repository?''  Users don't set up
%  repositories, they only check out code (on one system), and use SF
%  to mirror this.}
The code base can consist of components that are hosted in many
different repositories.  \simfactory\ will then replicate this source
tree structure from the home system to remote systems.

Compared to a more ``traditional'' development model, where one checks
out the source code on multiple systems, this has several advantages.
First, all authoritative source code versions are located on a single
system where they can be easily compared, backed up, etc.  Second,
source trees on different systems cannot diverge accidentally since
\simfactory's mirroring command can keep them up to date.  In
addition, our build environment (see section \ref{sec:configuring} below)
automatically tags and captures source trees as executables are built,
so that one can recreate the source tree that was used to build a
certain executable.

Duplicating a source tree to a remote system requires knowledge about
remote access methods, authentication, directory names, etc.;
\simfactory\ accomplishes this goal by facilitating remote access and
authentication with the remote resource using a pre-configured
authentication mechanism such as \emph{ssh} or the \emph{Globus
  Toolkit}.  If the remote system is not world-accessible, \simfactory\ uses
 ``trampolines'' (one or more intermediate, authorized machines) to
  complete a trusted authentication chain.
Building on these mechanisms, \simfactory\ performs the
actual synchronization using \emph{rsync} or a comparable mechanism.
It should be noted that this mechanism requires typically only a few
minutes with a cold file cache, and under ten seconds with a hot file
cache.\footnote{This assumes that there are only few modifications to
  the source tree, as is usual during a edit-compile-test cycle.
  These times depend greatly on the file system. A cold cache refers to the first 
  time these files are accessed in the edit-compile-test cycle, which requires the 
  files be read from disk, and a hot cache
  refers to the files being in memory in the file system cache.}
We show example timing results in figure \ref{tab:synctimings}.

\begin{figure*}
  % ES: I make this a figure instead of a table, since table captions
  % use all-upper-case are are unreadable.
  \centering
  \begin{tabular}{| l l | r r | r r |}
    \hline
    Machine	& Location	& full/cold	& full/hot 	& update/cold	& update/hot	\\ \hline
    numrel10	& CCT (local) 	&        52 	&       40	&           5 	&          5	\\
    Eric	& LONI 		&        57 	&       40 	&           5 	&          4	\\
    Queen Bee	& LONI 		&        64 	&       64 	&           7 	&          7	\\
    Kraken	& NICS 		&      2277	&     1955 	&          14 	&         15	\\
    Ranger	& TACC 		&       112 	&      106 	&          23 	&         16	\\ \hline
    % eric.loni.org* 		& 0m57.098s 	& 0m40.286s 	& 0m5.219s 	& 0m3.857s	\\
    % kraken.nics.tennessee.edu	& 37m56.535s 	& 32m34.913s 	& 0m14.117s 	& 0m14.787s	\\
    % ranger.tacc.utexas.edu 	& 1m52.326s 	& 1m46.217s 	& 0m22.672s 	& 0m15.957s	\\
    % queenbee.loni.org*	& 1m3.645s 	& 1m3.777s 	& 0m6.811s 	& 0m6.602s	\\
    % numrel10.cct.lsu.edu** 	& 0m52.064s 	& 0m40.044s	& 0m5.349s 	& 0m4.619s	\\ \hline
  \end{tabular}
  \caption{Source-tree synchronization timings.
    (All times in seconds, rounded to the nearest second.)
    This measures synchronizing the complete Einstein Toolkit
    \cite{einsteintoolkitweb} source
    tree from a machine at the CCT (LSU) to the specified machine.
    This source tree consists of about 50k files with about 450 MByte.
    (``full'' copies the tree for the first time, ``update'' has the
    source tree already present on the destination; ``cold'' and
    ``hot'' describe the state file system cache.)
    In all cases, updating an existing source tree takes only a few
    seconds.
    The initial copy on Kraken takes an inordinate amount of time;
    this is a property of Kraken's file system and is also
    seen independent of the Simulation Factory.}
  \label{tab:synctimings}
  \negfigurespace
\end{figure*}

This synchronization replicates not only source code but can also
replicate parameter files and input files

\subsection{Configuring and Building}
\label{sec:configuring}

Before running simulation one has to configure and build an executable
from the source tree.  These steps typically cannot be performed
automatically and require user input; often, a complex set of
inter-dependent configuration variables has to be set.  In fact, this
step is often so complex that only experts are able to configure and
build an application on a particular system even if the applications
themselves are portable.

There are two kinds of configuration settings for an application.
Some configuration settings select features of the application (e.g.\
whether a certain component is to be included, or what level of
optimization should be used).  Other configuration settings are
determined by the host system and have to be chosen ``just right'' to
make everything work (e.g.\ compiler version, compiler flags, paths to
external libraries, etc.)

Tools that are commonly used to help
configuring and building applications are either not available on or
not applicable to HPC systems.  System-level tools such as
\texttt{rpm}~\cite{rpmweb}
or \texttt{apt}~\cite{aptweb} are not
provided on HPC systems, and cannot be installed by end-users.
User-level tools such as \texttt{autoconf}~\cite{autoconfweb} do not work well because (a)
existing software is often installed in non-standard locations, (b) several
fundamentally different versions of a particular package may be
available, requiring a user choice (e.g.\ between Intel and PGI
compilers), and (c) testing whether a feature
is present or not may require executing an application which may
require using a batch system to submit a job with a turn-around
times measure in hours.  (Thus autoconf may require days to
complete.)  Finally, autoconf's approach of simply trying out whether a
feature is available may be considered abuse of the system and be
disallowed by usage policies.

% {ES: Review 4 complains that the autoconf discussion above is not
%   completely true.  However, it speaks about cross-compiling (which we
%   don't quite address), and asks how SF overcomes the autoconf
%   problems described above (with the mdb!).}
% {ES: I am going to ignore this comment; I believe we describe the
%   mdb well enough below.}

% {ES: Review 4 suggest to compare to CMake as well.}
\texttt{CMake}~\cite{cmakeweb} has
features similar to \texttt{autoconf}; there is no provision for a
database that would store machine-specific configuration details.

Most HPC systems allow user-level configuration tools such as SoftEnv~\cite{softenvweb}
or Environment Modules~\cite{envmodweb}.  These packages allow users
to choose between several installed software versions (e.g.\ different
compilers), and (if set up properly) will ensure that all enabled
software packages are compatible with each other.  Unfortunately,
SoftEnv settings are global to a user's account, so if a user builds
an executable, and then switches his/her SoftEnv settings while this
executable is waiting in a queue, the job may crash when it finally
starts since important libraries may not be available any more.
Environment Modules remedy this flaw but they still do not provide
sufficient information to (a) choose a particular version of a module,
or (b) select good compiler options for this version.  It is also not
possible to automatically determine what library a particular module
provides, as only a module's name is available -- the BLAS~\cite{blasweb} library may
e.g.\ be provided by modules called \texttt{acml}~\cite{acmlweb}, \texttt{atlas}~\cite{atlasweb},
\texttt{essl}~\cite{esslweb}, \texttt{gotoblas}~\cite{gotoblasweb}, or \texttt{mkl}~\cite{mklweb}, (and none of which
provide the Netlib reference implementation \cite{blasweb}).
Similar ambiguities exist with choosing a compiler or with choosing
an elliptic solver if the application supports several interfaces
(e.g.\ both PETSc and Hypre).  And in the end, there may still be
incompatibilities due to errors or mis-interpretations between any of
these or any of these and the application, which may only be
detectable as crashes at run time.

What this all amounts to is that one
has to carefully study a system's documentation and perform careful
experiments to find good, working configuration options.

\simfactory\ stores the complete configuration information for every
machine (for a given application) in its machine database.  This
allows everybody (including new users) to quickly build the
application on every supported system.  This configuration information
is thus, in its entirety, ``blessed'' by its corresponding maintainer,
providing a service to the community.  This approach is in contrast to
a design where the simulation factory would pick up bits and pieces of
configuration information automatically from the host system and
would then combine the configuration information by itself; this
approach lacks the ``seal of approval'' that exists in the current
approach.  Given the complexity of identifying a correct set of
configuration information for a particular system, where even small
details can uncover compiler errors or lead to inconsistencies between
different libraries, such a seal of approval is crucially important.

The disadvantage is that this approach does not scale well -- it
requires one maintainer per application per supported system and the
Cactus configuration information with about 30 supported production
system is reaching this limit.
% {ES: Review 6 requests us to offer solutions to this
%   non-scalability.  For example, what about having one mdb for each
%   user community?}
We suggest to address this in two ways: In the short term, we assume
that each user community will maintain their own machine databases,
limiting the number of systems that need to be supported in each of
these.  In the long term, we expect resource providers to maintain
more and more details about their systems in formal descriptions in a
reliable manner, so that parts of the machine database can be
determined automatically.

The Cactus framework supports building multiple executables from the
same source trees.  Each of these executables is called a
\emph{configuration}, since they differ in the configuration settings
that were used to create them.  These configuration settings include a
\emph{thorn list} (list of components that should be included into the
configuration), and a set of high-level build options such as
\emph{debug}, \emph{optimise}, or \emph{profile}.  These
configurations have different names, allowing the user to specify
which configuration to use when performing a simulation.

In many cases, a single \simfactory\ command suffices to create a Cactus
executable from a given thorn list and the very same command works on
any production machine that a research group may be using.  This is a
substantial simplification over the previous state of affairs and
allows every Cactus user to add or modify code and then re-build the
configuration with only minimal effort.

\subsection{Managing Simulations}
\label{sec:managing}

Once a configuration has been created, \simfactory\ provides
functionality for performing simulations, either by submitting and
managing jobs via a job queueing system, or by directly running the
executable.

\begin{figure}
  \centerline{\includegraphics[width=\linewidth]{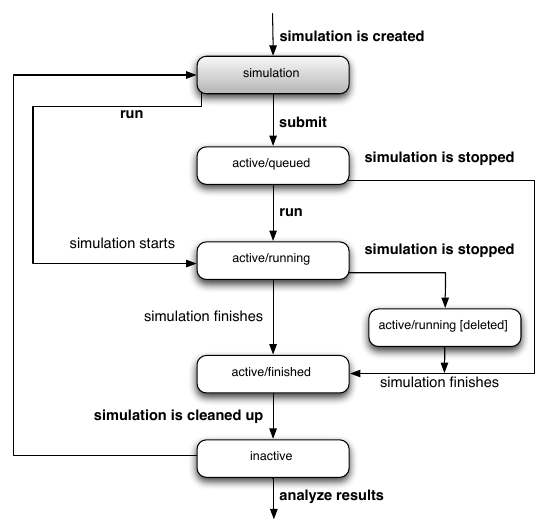}}
  \caption{Simulation Workflow.  This shows the life cycle of a
    simulation, starting from its creating, proceeding potentially
    through several restarts, until it is finished. The bold text directly 
    corresponds to \simfactory\ commands.}
  \label{fig:simworkflow}
  \negfigurespace
\end{figure}

The simulation factory follows a high-level abstraction for performing
simulations.  In this abstraction, performing a simulation consists of
many more operations than just running the executable.  In particular:
\begin{itemize}
\item The first step is to \emph{create a simulation}, which captures
  an executable, a parameter file, any input files there may be, as
  well as any other parameters that determine the physical result of
  the simulation.  This excludes any incidental parameters (see
  below), and does not actually start the simulation.
\item To make progress with the simulation, one either submits
  or runs a \emph{restart} (roughly equivalent to a ``job'' in a queueing system).  
  This requires choosing incidental parameters, such as the number 
  of processes, wall time limit, allocation/queue names, etc.  At most 
  one restart can be active at one time. 
\item After the restart finishes, it is \emph{cleaned up}, which may
  consist of minor actions such as correcting file permissions, or
  deleting unnecessary or broken files, or may e.g.\ automatically
  archive simulation results.
\item A simulation may consist of arbitrarily many restarts.
  Presumably, each restart continues where the previous restart left
  off.  (This is necessary since queue time limits on HPC systems are
  often much too short to complete a simulation.)
\end{itemize}
Figure \ref{fig:simworkflow} shows the life cycle of a simulation and
its restarts graphically.

This high-level model for performing a simulation differs in certain
crucial respects from a ``naive'' way of doing so.  First, it
distinguishes clearly between those parameters that influence the
scientific result and those that are only incidental.  Second, the
simulation factory explicitly captures all parameters and necessary
input (by copying) so that one cannot accidentally modify these
parameters while a simulation is running, which can be a time scale of
many weeks.  Third, it introduces the notion of a clean-up step after
each restart has run, providing a hook for additional actions, such as
automatically archiving results.

Finally, the simulation factory performs these details automatically,
consulting the machine database to be able to map these details onto a
particular system.  By performing these actions automatically, (a)
each simulation (within a research group) is performed in a consistent
manner, allowing others to understand the directory layout and access
the results, (b) users are relieved of performing many tedious
low-level tasks manually, and (c) many errors and accidental omissions
are avoided.  By implementing best practices in the simulation
factory, the often unwritten experience of users is captured,
resulting in simulations that are ``naturally'' performed in the
``best possible'' manner, including e.g.\ capturing all necessary
provenance information to make simulations truly repeatable.

In detail, the simulation factory performs the following actions:
\begin{itemize}
\item create and populate a directory structure for the simulation;
\item copy all executables, parameter files, and input files to save
  them from accidental modification (ensuring to avoid superfluous
  copies to save disk space);
\item manage checkpoint files, and/or ensure the correct checkpoint
  files are used when restarting;
\item allow \emph{presubmitting} (or \emph{chaining}) of restarts,
  i.e.\ submitting several restarts at once that will then
  automatically be executed sequentially (if and as allowed by the
  queueing system and system policies);
\item allow to run executables directly or interactively, potentially
  using ``interactive queued jobs'';
\item use and/or create unique identifiers for simulations and
  restarts, so that output files can be tagged pointing back to the
  exact simulation;
\item automatically archive simulations and restarts, since files on
  HPC systems are often automatically deleted after some time;
\item clean up in various minor but convenient ways, e.g.\ correct
  file permissions or delete unnecessary output files.
\end{itemize}

\subsection{Information and Status Inquiries; Provenance}

In addition to the actions supporting source tree management,
configuring/building, and managing simulations, \simfactory\ offers
commands to display information and inquire about the status of
simulations and restarts.  This includes displaying \mdb\ entries and
configuration information.
This allows the user to monitor simulations,
know when they are completed, and check and restart them if they are
interrupted.  This also provides access to the
list of the HPC resources that have been pre-configured, allowing the
user to know which HPC resources they can begin using immediately.

% {ES: Review 1 suggests to speak about provenance.  We should
%   mention what we do here, e.g.\ creating unique IDs and keeping log
%   files for user actions and his/her parameter choices.  SF also knows
%   about Formaline and shows Formaline output to the user.  In future
%   work, speak about SF archiving these IDs, log files, and Formaline
%   output.}

In performing the actions described above, \simfactory\ keeps a log
trail associated with each simulation, describing all steps that led
to a particular simulation state and the order in which they were
taken. \simfactory\ also keeps safe copies of all executables and
input files that were used for a particular simulation so that
changes to these files outside of \simfactory's control do neither
affect currently ongoing simulations nor confuse the log trail.

Furthermore, \simfactory\ generates and manages unique identifiers for
each simulation and restart and passes these and other provenance
information on to the application which can then e.g.\ tag output
files and images. \simfactory\ also understands provenance information
generated by the application, which is stored together with
application output and which is displayed to the user together with
other application output.

\subsection{Machine Database}
\label{sec:mdb}

The main idea behind \simfactory\ is to provide a resource
agnostic tool that is able to take full advantage of the various HPC
resources available to researchers.  To separate
\simfactory\ from the resource on which it is used, we employ a
\emph{machine database}, which is a collaborative database compiled by
\simfactory\ maintainers, researchers, and other users that stores all machine dependent information.
This database is of crucial importance to \simfactory's design (see
section \ref{sec:configuring} above).

Each machine database entry is broken out into four groups:

\begin{itemize}
\item \emph{Machine description:} information about the machine itself, used
  for logging/informational purposes;
\item \emph{Machine access:} parameters defining remote access and
  management of the resource, e.g.\ defining how to log in or to
  synchronise files and directories;
\item \emph{Source tree management:} parameters for defining a good
  directory for the source tree, the default make command, and what
  the default configuration options are;
\item \emph{Simulation management:} parameters containing information on how
  to submit, manage, and run simulations using the resource's specific
  job queueing system.
\end{itemize}

This database provides the user the ability to run \simfactory\ on any
known resource.  We use the word ``known'' because the \mdb\ is a
collection of pre-defined settings that represent the knowledge
\simfactory\ must have in order to configure, build, and run
simulations.  The \mdb\ provides the ability to deploy simulations quickly and
efficiently.

%\todo{ES: Review 4 points out that the begin--to--end workflow for the
%  end user is not described.  Could add another subsection here, or
%  describe it earlier near figure \ref{fig:lifecycle}. Michael, do you
%  have an idea here?}

%%%%%%%%%%%%%%%%%%%%%%%%%%%%%%%%%%%%%%%%%%%%%%%%%%%%%%%%%%%%%%%%%%%%%%%%%%%%%%%%

\section{Implementation Issues}
\label{sec:implementation}

Below we discuss some interesting challenges that had to be addressed
in the design and implementation of \simfactory.  Specifically, we
will focus on the evolution of \simfactory\ to its current state, the method used by 
\simfactory\ to become machine agnostic, the abstraction we use to communicate
\simfactory's command line options between its modules, the restart
abstraction, which represents how \simfactory\ encapsulates a specific
stage of a simulation and finally how \simfactory\ represents
presubmitted jobs internally, which involves creating and submitting
multiple jobs to a job queuing system to handle simulation wall times
greater than the maximum allowed wall time for a single job.

\subsection{History}

\simfactory\ was originally designed for SC~2006 to simplify executing
sets of benchmarks on many HPC systems, and was implemented
as a monolithic Perl script
centered around the \mdb.  The \mdb\ was implemented as Perl hash table
initialized via executable Perl code.  The three different tasks
(source code management, configuring and building, simulation
management) existed from fairly early on, as did the distinction
between a simulation and a restart.  Advanced features such as
presubmission was ``tacked on'' later and didn't quite fit the
original abstract model of a simulation, and neither did running a
restart without submitting it to a queue.  In addition, after growing
a user base, maintaining the \mdb\ and its necessary local
modifications in executable Perl code was not attractive any more.

While updating the abstract model of a simulation, we decided to
restructure \simfactory's internal implementation and to switch to
Python as implementation language because several of the main users
and contributors requested so.\footnote{The main reasons cited for
  preferring Python was the cleaner syntax and the more elegant modern
  facilities to support large-scale programming, such as classes and
  modules.  One reason cited against Python was that it does not
  perform a static type checking at startup, so that uninitialized
  variables are only detected at run time.}
One of the main features of the
\simfactory\ is that it can run everywhere without being installed
(just downloading or copying must suffice), which essentially leaves
only Bash, Perl, or Python as implementation languages.  We found that
most HPC systems run Python 2.3 or later -- this also nicely
demonstrates the extremely conservative time scale on which HPC
systems install or update software.\footnote{Python 2.3 was released
  in 2003; we write this paper in 2010.}

% {ES: Review 4 states that simply rewriting in Python is not good
%   enough for a paper.  Need to point out that this is the first paper,
%   and don't emphasise the rewriting.  Maybe simply speak of two
%   implementations.}
% {ES: I am ignoring this comment.}

We rewrote \simfactory\ in Python as an object-oriented, module-based 
application.
\simfactory\ consists of four separate modules, corresponding to the
features described in the subsections of section
\ref{sec:simfactorybasic} above.
%~\ref{simbinlayout}
Each module can either be called directly by the user, or can be
dispatched as an imported module by a (future) GUI application.
\simfactory\ currently includes one command line UI application called
\texttt{sim}, maintaining command-line compatibility with the previous
version while also providing access to new functionality.  Each of
these four main modules relies heavily on the \mdb\ to provide
information about the HPC resource on which it is executing.

\subsection{Machine Database Details}

\subsubsection{File Format}

As described in section \ref{sec:mdb} above, the machine database
describes properties of each machine so that \simfactory\ can provide
a resource-agnostic user interface.

The HPC user community dislikes non-human-readable file formats for
non-performance critical information.  We therefore use the \ini\ file
format~\cite{wikiini},
extended to support multi-line block text entries similar to ``here''
documents found in Bash and Perl.  Figure \ref{fig:mdbexample}
shows a partial entry for a particular machine, and figure
\ref{fig:iniblock} demonstrates a multi-line entry.  Multi-line block
entries use \blockstart\ followed by any sequence of characters and
then a newline to indicate the start of a block.  To end the block,
the same sequence of characters following \blockstart\ must be
repeated on a line by themselves.

We chose this file format because it is both readable and portable.
The \ini\ file format is platform agnostic, clear-text, and easily
understood, created, and edited because of its simple syntax.

\begin{figure}
{\small
\begin{Verbatim}[frame=single, framerule=0.3mm]
# a block of text 

# Official Cactus entries
sync-sources   = <<EOT
CONTRIBUTORS
COPYRIGHT
Makefile
arrangements
src
lib
EOT

\end{Verbatim}
}
\caption{Example of a multi-line block of text in the machine database, listing multiple directory names for the key \texttt{sync-sources}.}
\label{fig:iniblock}
\negfigurespace
\end{figure}

\subsubsection{Implementation}

The \mdb\ is comprised of a set of entries with predefined keys,
separated into sections with unique names for each machine, as shown
in figure~\ref{fig:mdbexample}.

\begin{figure}
{\small
\begin{Verbatim}[frame=single, framerule=0.3mm]

[queenbee]

# Machine description
nickname        = queenbee
name            = Queen Bee
location        = LONI, LSU
description     = The large LONI Linux clu...
webpage         = http://www.loni.org/...
status          = production

# Access to this machine
user            = YOUR_LOGIN
email           = YOUR@EMAIL.ADDRESS
hostname        = qb4.loni.org
rsynccmd        = /home/eschnett/rsync-3.0...
sshcmd          = ssh
localsshsetup   = :
sshsetup        = :
aliaspattern    = ^qb[0-9](\.loni\.org)?$

# Source tree management
sourcebasedir   = /home/@USER@
...
\end{Verbatim}
\iffalse
$
\fi
}
\caption{Example partial Machine Database entry in the human-readable
  \ini\ file format.  This is taken from the entry describing Queen
  Bee, a shared LONI/TeraGrid system in Baton Rouge, LA\@.}
\label{fig:mdbexample}
\negfigurespace
\end{figure}

% \begin{description}
% \item[Machine description:] Information about the machine itself, and
%   is used for logging/information purposes.  [\emph{nickname, name,
%     location, description, webpage, status}]
% \item[Machine access:] Parameters defining remote access and
%   management of the resource.  These keys define how to login, whether
%   or not it is necessary to use a secondary resource in order to
%   access this resource, both for file copy and for remote access
%   purposes, and also how to synchronise files and directories.
%   [\emph{user, email, hostname, iomachine, trampoline, rsynccmd,
%     rsyncopts, sshcmd, sshopts, localsshsetup, sshsetup,
%     aliaspattern}]
% \item[Source tree management:] Parameters for defining a good
%   directory for the source tree, the default make command, and what
%   the default option list, component list, and script files are.
%   [\emph{sourcebasedir, optionlist, thornlist, scriptfile, make}]
% \item[Simulation management:] Parameters containing information on how
%   to submit, manage, and run simulations using the resource's specific
%   job queueing system.  [\emph{basedir, quota, cup, cpufreq,
%     flop/cycle, ppn, spn, mpn, max-num-threads, num-threads, memory,
%     nodes, min-ppn, allocation, queue, maxwalltime, maxqueueslots,
%     submit, run, run2, getstatus, stop, submitpattern, statuspattern,
%     queuedpattern, runningpattern, scratchdir, exechost,
%     exechostpattern, stdout, stderr, stdout-follow}]
% \end{description}

We anticipate that users will have to maintain local modifications of
the \mdb, typically to define their user names and allocations on
these system.  To facilitate this, \simfactory\ provides the user with
the ability define global defaults in a \texttt{[default]} section, as
shown in figure~\ref{fig:defexample}.  Settings defined here propagate to
all \mdb\ entries.  Users can also provide new \mdb\ entries,
describing e.g.\ private systems such as personal workstations or
notebooks, and can overwrite specific keys in the \mdb.
Figure~\ref{fig:mdboverride} demonstrates how to define a different user
name and thorn list, which will also override the value in the
\texttt{[default]} section, for a pre-defined \mdb\ entry.

\begin{figure}
  \centerline{\includegraphics[width=0.75\linewidth]{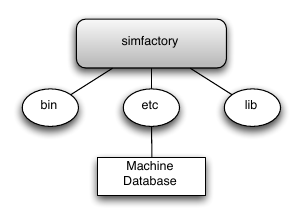}}
  \caption{Simulation Factory db file structure.}
  \label{fig:dbfilestruct}
  \negfigurespace
\end{figure}

\begin{figure}
{\small
\begin{Verbatim}[frame=single, framerule=0.3mm]

# override some mdb entries
[queenbee]
user            = mthomas
thornlist       = einstein-bassi.th

[is]
user            = mthomas
thornlist       = einstein-numrel-intel.th

\end{Verbatim}
}
\caption{Machine database syntax: Specifying a different user name for
  certain machines in the machine database.}
\label{fig:mdboverride}
\negfigurespace
\end{figure}

\begin{figure}
{\small
\begin{Verbatim}[frame=single, framerule=0.3mm]

# Defaults for all machines
[default]
user        = mwt
email       = mthomas@cct.lsu.edu
allocation  = loni_cactus04

\end{Verbatim}
}
\caption{Machine database syntax: Defining default values to override
  entries in the machine database.}
\label{fig:defexample}
\negfigurespace
\end{figure}

The \mdb\ comes pre-configured and is maintained by the developers
and maintainers of \simfactory. The last database, the \udb, is where
the user makes their edits, adds their local workstation if necessary,
and defines necessary default values that will propagate throughout
each of the three databases.

The databases are located in the \simfactory\ source tree inside the
\texttt{etc} folder, as seen in Figure~\ref{fig:dbfilestruct}.

\subsection{Macros}

Machine database entries, option lists, script files, and other files
that are processed by the \simfactory\ can contain macros that are
replaced at run time.  For example, every instance of \texttt{@USER@}
found in the \texttt{[queenbee]} \mdb\ entry (see figure
\ref{fig:mdbexample}) will be replaced by the username
\texttt{mthomas}, as specified in figure \ref{fig:mdboverride}.  This
allows the user the flexibility to set the user name once and use it
in a generic manner for any other user-name dependent parameters
needed by \simfactory.

In addition to a set of predefined macros that correspond to \mdb\
entries or values gathered from the system (such as e.g.\ a simulation job
id), one can also define additional, arbitrary macros via command line
options to \simfactory.

\subsection{Restarts}

One unique challenge \simfactory\ addresses is how to manage multiple
stages of a single simulation, whether the stage represents a fresh
start or a recovery from a previously interrupted simulation.  To
address this, \simfactory\ contains an abstraction called
\emph{restart}.  Figure~\ref{fig:simworkflow} in section
\ref{sec:managing} above shows the typical life of a \simfactory\
restart.  A restart encapsulates several key operations:
\begin{itemize}

% rewritten for clarity. Is this any better?
\item \emph{create:} Before a simulation can be submitted or executed, it must be created. 
  This process initializes any required configuration parameters, directories, and any other
  infrastructure necessary to facilitate submission or execution.
\item \emph{submit:} The submit operation submits the simulation to the
  queueing system of the host system.  A shell script file 
  is the command that is sent for submission.  This
  script contains the necessary options to initialize the queueing
  system correctly, such as choosing the correct wall time, and then
  commands calling the \simfactory\ \texttt{run} operation.  This
  submit operation does not contain any execution logic.
\item \emph{run:} The run operation is where all execution logic lives.  The
  run command is capable of launching simulations that have either
  been submitted in a queuing system, or are executed directly,
  without the use of a queueing system.  The ability to execute a
  simulation directly is necessary for running small simulations on a
  local workstation or notebook.  The run operation uses the
  \texttt{mpirun} command specified in the \mdb.  This command can be
  any shell command, such as the actual \texttt{mpirun} command to
  launch a job using the system's MPI implementation, or just the
  simulation binary itself, skipping the use of the MPI altogether.
\end{itemize}

\subsection{Presubmission}

One of the most important challenges \simfactory\ has to address is
the need to execute simulations automatically beyond the maximum
execution time allowed on the HPC resource.  To facilitate this,
\simfactory\ determines, based upon the specified and maximum allowed
wall time, how many submissions to the queueing system are necessary.
\simfactory\ then submits the job this many times and then sets up
the necessary daisy-chain to make the simulations launch in the
correct order.  A simulation checkpointing system, which allows a
simulation to resume where it has left off, is necessary for
presubmission.

\section{Future Plans}
\label{sec:future}

We plan to extend the feature set provided by \simfactory\ to include
managing simulation output, i.e.\ often large output files that are
``left behind'' after simulations have finished.  Managing simulation
output also depends on low-level machine characteristics, such as
after how much time it will be deleted automatically, what commands
have to be used to access it efficiently, and what long-term storage
mechanisms can be reached from the simulation machine.  Other
challenges include ensuring that simulation output files and
post-processing results remain ``together'', so that one can reliably
trace back to the root of the simulation e.g.\ from a figure found in
a publication or an image found on a web site.  This topic involves
repeatability of numerical calculations and provenance of numerical
data, and the \simfactory\ is an ideal vehicle to implement the
corresponding low-level mechanism to spare the user these details.

Another possible extension of \simfactory\ is to support simulations
running across multiple sites or off-loading certain ancillary
(analysis) tasks to secondary machines to reduce the load on the
primary machine.  In the past, several of the issues relating to
identifying such tasks in the Cactus framework were addressed
\cite{Allen01a, Lanfer01a}, but reliably being able to mirror the
source code onto remote systems and reliably starting new simulations
there remained an unsolved problem.  This will be an ideal application
case for the \simfactory.

While \simfactory\ is currently targeting the Cactus framework, it is
clear that the issues it addresses and the solution it offers are
relevant for many other applications as well.  We intend to abstract
out all application-specific bits into an \emph{application database}
so that \simfactory\ can support other applications as well, as long
as these require any of the same basic features (access to remote
systems, source code management, simulations on HPC systems).

Other simulation packages (e.g.\ Enzo, ADCIRC, NAMD, etc\@.) face very
similar issues. If a package is used only as ``black box'' on an HPC
system, without making modifications to its source code, one can use a
pre-installed copy of the executable which often mitigates these
problems. However, it is the nature of graduate student research that
new methods and new algorithms be tried out and these will then have
to face the issues. The basic workflow (build and test on local
system, build and test on HPC system, submit and manage a set of jobs,
handle checkpointing and restarts) is virtually identical for most
simulation packages.
We have recently
begun to investigate \simfactory\ support for ADCIRC~\cite{adcircweb}~\cite{Swathi:TeraGrid:2010},
%\todo{cite Swathi's TG'10 poster},
which is used to model storm surges caused
e.g.\ by hurricanes in the Gulf of Mexico. 

% ES: Review 6 complains that we claim generality (not just
%   Cactus) which we need to back up by examples.  For example, describe
%   similarities between workflows with Cactus and other simulation
%   packages.}
% ES: Should also speak about ADCIRC, (NAMD, GEOS5, LAMMPS, WRF),
%   including usability.}

While the \simfactory\ provides the infrastructure (middleware) for
the features it offers, it also provides only command-line interfaces.
Graphical User Interfaces (GUIs) would provide several obvious
advantages; first, guiding the user to possible next steps for each
state (making \simfactory\ easier to use for beginners), and second,
providing visual feedback about the current state and keeping it
updated automatically, without requiring a user command and the lag
time thereafter.  The (command line) user interface logic is separate
from the application logic, which will allow us or others to offer a
GUI, based e.g.\ on QT~\cite{qtweb} or GTK~\cite{gtkweb}, or via a web
service.  We are also currently implementing a TeraGrid Gateway~\cite{teragridgatewayweb}
for numerical relativity that will use the Simulation
Factory.

Since being accepted, this work has also been presented as a poster at
TeraGrid '10~\cite{Thomas:TeraGrid:2010}.

\simfactory\ is released under an open source license and is freely
available as part of the Einstein Toolkit, as described on
\url{http://einsteintoolkit.org}.

\section*{Acknowledgments}

We thank Ian Hinder for his suggestions and contributions toward the
current state of the Simulation Factory, in particular for mercilessly
simplifying the user interface and suggesting and driving the
implementation of presubmitting.  We also thank all users who have contributed or 
updated machine database entries.  We especially thank Gabrielle Allen for her insightful 
and helpful comments while writing this paper.

We acknowledge support from the awards NSF OCI 0721915 \emph{Cactus
  Tools for Application Level Performance and Correctness Analysis
  (Alpaca)}, NSF PHY 0701566 \emph{XiRel, A Next Generation
  Infrastructure for Numerical Relativity}, NSF PHY 0904015
\emph{Community Infrastructure for General Relativistic MHD (CIGR)},
NSF OCI 0932251 \emph{TeraGrid Extension: Bridging to XD},
and the REU project \emph{Interdisciplinary Research Experience in
  Computational Sciences} via NSF award OCI 1005165 and NSF/Louisiana
Board of Regents award ESP 0701491.
MT gratefully acknowledges financial support from the TeraGrid to
attend the TeraGrid '10 conference.
We accessed HPC resources on the
TeraGrid via allocation TG-MCA02N014, at NERSC supported by DOE
contract DE-AC02-05CH11231, and on LONI under the allocations
loni\_cactus and loni\_numrel.

\bibliographystyle{IEEEtran}

\bibliography{publications-schnetter,publications-thomas,references,simfactory}

\end{document}